\documentclass[aps,prb,showpacs,twocolumn,amsmath,amssymb,
superscriptaddress,letterpaper]{revtex4-1}
\usepackage{mathrsfs}
\usepackage{graphicx}
\usepackage{graphics}
\usepackage{float}
\usepackage{subfigure}
\usepackage{bm}
\usepackage{dcolumn}
\usepackage{amsmath,bm}
\newcommand{\overbar}[1]{\mkern 1.5mu\overline{\mkern-1.5mu#1\mkern-1.5mu}\mkern 1.5mu}

\usepackage{color}   
\usepackage{hyperref}
\hypersetup{
    colorlinks=true, 
    linktoc=all,     
    linkcolor=blue,  
    citecolor=black
}

\begin{document}
\title{ Higgs amplitude mode in massless Dirac fermion systems}
\author{Ming Lu}
\affiliation{International Center for Quantum Materials and School of Physics, Peking University, Beijing 100871, China}
\affiliation{Collaborative Innovation Center of Quantum Matter Beijing 100871, China}

\author{Haiwen Liu}
\affiliation{International Center for Quantum Materials and School of Physics, Peking University, Beijing 100871, China}
\affiliation{Collaborative Innovation Center of Quantum Matter Beijing 100871, China}

\author{Pei Wang}
\affiliation{International Center for Quantum Materials and School of Physics, Peking University, Beijing 100871, China}
\affiliation{Department of Physics, Zhejiang Normal University, Jinhua 321004, China}

\author{X. C. Xie}
\affiliation{International Center for Quantum Materials and School of Physics, Peking University, Beijing 100871, China}
\affiliation{Collaborative Innovation Center of Quantum Matter Beijing 100871, China}
\date{\today}

\begin{abstract}
The Higgs amplitude mode in superconductors is the condensed matter analogy of Higgs bosons in particle physics. We investigate the time evolution of Higgs amplitude mode in massless Dirac systems, induced by a weak quench of an attractive interaction. We find that the Higgs amplitude mode in the half-filling honeycomb lattice has a logarithmic decaying behaviour, qualitatively different from the $1/\sqrt{t}$ decay in the normal superconductors. Our study is also extended to the doped cases in honeycomb lattice. As for the 3D Dirac semimetal at half filling, we obtain an undamped oscillation of the amplitude mode. Our finding is not only an important supplement to the previous theoretical studies on normal fermion systems, but also provide an experimental signature to characterize the superconductivity in 2D or 3D Dirac systems. 
 
\end{abstract}


\maketitle

\section{INTRODUCTION}
A conventional superconductor can be described by a charged complex order parameter $\Delta(r,t)=|\Delta(r,t)|e^{i\phi(x,t)}$. Its collective fluctuations around equilibrium including the oscillations of the phase and amplitude\cite{Varma2015}. The phase mode, being coupled to the electromagnetic field, moves to plasma frequency of the metal as a manifestation of Anderson-Higgs mechanism\cite{Anderson1963,Higgs1964,Greiter2004}. The amplitude mode oscillates with the angular frequency $2|\Delta_0|$, analogous to the ``vibration" of the longitudinal component of Higgs field in particle physics\cite{Pashkin2014}. In this sense, the amplitude mode in superconductor is sometimes also called Higgs mode or Higgs amplitude mode in the literature\cite{Varma2015,Pashkin2014,Matsunaga2013,
Matsunaga2014,Sherman2015,Tsuji2014}.

Higgs amplitude mode in superconductors, although theoretically predicted many years ago\cite{Varma1982}, has only been directly observed recently by the time-resolved Teraherz (THz) pump-probe technique in a clean superconducting film\cite{Matsunaga2013,Matsunaga2014}, and by measuring the excess sub-gap optical conductance in disordered films near the superconductor-insulator phase transition\cite{Sherman2015}. The time evolution of the Higgs mode in the collisionless, dissipationless regime was studied intensely. It was revealed that the Higgs mode oscillates at a frequency of $2\Delta_\infty$ with a $1/\sqrt{t}$ decaying property in the weak coupling limit, where $\Delta_\infty$ is the asymptotic value of superconducting gap
\cite{Volkov1973,Levitov2006,Yuzbashyan2006Linear,Tsuji2014}. 
However, previous works all assume that the density of states (DOS) near the Fermi level is almost a constant within the Debye cut-off energy $\omega_D$. This assumption obviously fails for honeycomb lattice or Dirac semimetals at half filling. Their DOS is either linear (2D) or quadratic (3D) at low energy, respectively, and vanishes at the Dirac point\cite{Wallace1947,Young2012}. Since superconductivity is strongly affected by the DOS near the Fermi level, it would be theoretically interesting to study the time evolution of Higgs mode in those systems. On the experimental side, the availability of the honeycomb optical lattice\cite{Esslinger2012} and the tunable attractive interaction by Feshbash resonance\cite{Esslinger2010} give a possible test ground for this study. Besides, the expected unique feature of the Higgs mode in superconducting Dirac semimetal can be used as an important experimental characterization to distinguish it from the normal superconductors\cite{Wangjian2015,Aggarwal2015}.

In this paper we study the quenched dynamics in the weak coupling limit by using the Anderson pseudo-spin formalism\cite{Anderson1958}. We find that the Higgs mode has a log-decay behaviour in the half-filling honeycomb lattice. To understand this behaviour, we further study the pseudo-spins' phase dynamics, and analytically solve the linearized equations of motion\cite{Volkov1973,Yuzbashyan2006Linear,Tsuji2014}. The doped cases is also studied numerically. In the low doping limit, a double-frequency feature is found. The larger frequency increases noticeably and its peak broadens with the doping level. In the high doping limit, we are back to the $1/\sqrt{t}$ decaying property, as in a normal superconductor. When considering the 3D Dirac semimetal at neutral point, we find that the Higgs mode exhibits an undamped oscillation, with all the pseudo-spins precess synchronizely.

\section{Model and formalism}
We start by considering the negative-U Hubbard model on honeycomb lattice:
\begin{equation}
\hat{H}=-\sum_{<ij>,\sigma}\hat{a}_{i\sigma}^\dagger b_{j\sigma}+h.c.-U\sum_{i}\hat{n}_{i\uparrow}\hat{n}_{i\downarrow}-
\mu\sum_{i\sigma}\hat{n}_{i\sigma}
\end{equation}
where $\hat{a}_i$ ($\hat{b}_i$) is the on-site annihilation operator on sub-lattice A (B); $\hat{n}_{i\sigma}$ is the number operator on lattice site $i$ with spin index $\sigma$; $\mu$ is chemical potential and $U$ is the the on-site attractive interaction. We choose the nearest-neighbour hopping as the energy unit throughout this paper. 

To study the dynamics, we 
write out the corresponding mean-field Hamiltonian in $\bf{k}$-space after a unitary transformation:  $\hat{a}_{\bf{k}\sigma}=\frac{1}{\sqrt{2}}(e^{i\theta_{\bf{k}}} \hat{c}_{\bf{k}\sigma}+\hat{d}_{\bf{k}\sigma}),\, \hat{b}_{\bf{k}\sigma}=\frac{1}{\sqrt{2}}(-\hat{c}_{\bf{k}\sigma}+e^{-i\theta_{\bf{k}}}\hat{d}_{\bf{k}\sigma})$:
\begin{align}
H_{MF} =& -\sum_{\bf{k}}(\mu-|\gamma_{\bf{k}}|) \hat{c}_{\bf{k}\sigma}^\dagger c_{\bf{k}\sigma}-\sum_{\bf{k}}(\mu+|\gamma_{\bf{k}}|) \hat{d}_{\bf{k}\sigma}^\dagger d_{\bf{k}\sigma}  \nonumber
\\
& - \Delta^*(t)\sum_{\bf{k}}\left(\hat{c}_{\bf{k}\uparrow}^\dagger c_{-\bf{k}\downarrow}^\dagger+\hat{d}_{\bf{k}\uparrow}^\dagger d_{-\bf{k}\downarrow}^\dagger\right)+h.c.
\end{align}
where $\hat{a}_{\bf{k}\sigma}$ ($\hat{b}_{\bf{k}\sigma}$) is the Fourier component of $\hat{a}_i$ ($\hat{b}_i$);  $e^{i\theta_{\bf{k}}}=\gamma_{\bf{k}}/|\gamma_{\bf{k}}|$ with $\gamma_{\bf{k}} = \sum_{\bf{k}} e^{i\bf{k}\cdot \bm{\delta}}$ and $\bm{\delta}$ being the three real space nearest-neighbour vectors; the time dependent order parameter $\Delta(t)=\frac{U}{N_c}\sum_{\bf{k}}\left<a_{\bf{k}\uparrow}^\dagger a_{-\bf{k}\downarrow}^\dagger\right>=\frac{U}{N_c}\sum_{\bf{k}}\left<b_{\bf{k}\uparrow}^\dagger b_{-\bf{k}\downarrow}^\dagger\right>$, in which $N_c$ is the number of unit cells and $\left<\cdots\right>$ denotes the time dependent quantum-mechanical expectation value.

We define two set of Anderson pseudo-spins: $\hat{\bm{S}}_{\bm{k}}^{(+)}=\frac{1}{2}\left(\hat{c}_{\bm{k}\uparrow}^\dagger,\,\hat{c}_{-\bm{k}\downarrow}\right)\bm{\sigma}
\binom{\hat{c}_{\bm{k}\uparrow}}{\hat{c}_{-\bm{k}\downarrow}^\dagger}$, $\hat{\bm{S}}_{\bm{k}}^{(-)}=\frac{1}{2}\left(\hat{d}_{\bm{k}\uparrow}^\dagger,\,\hat{d}_{-\bm{k}\downarrow}\right)\bm{\sigma}
\binom{\hat{d}_{\bm{k}\uparrow}}{\hat{d}_{-\bm{k}\downarrow}^\dagger}$, with their corresponding local fields $\bm{b}_{\bm{k}}^{(\pm)}(t)=\left( \Delta^{\mathrm{R}}(t),\,\Delta^\mathrm{I}(t),\,\mu\mp|\gamma_{\bf{k}}|\right)$. It is straightforward to check that the pseudo-spin operators satisfies the commutation relationship of the angular momentum (with $\hbar=1$). Using the above definition, the Hamiltonian can be written as the sum of the ``Zeeman energy" of pseudo-spins in their corresponding local fields:
\begin{align}\label{spin-H}
H_{MF}=-2\sum_{\mathbf{k},i=\pm} \bm{b}_{\bf{k}}^{(i)}\cdot \hat{\bm{S}}_{\bm{k}}^{(i)} 
\end{align}
From the Hamiltonian, we can get the equations of motion of pseudo-spins: $
\frac{\partial}{\partial t} \bm{S}_{\bm{k}}^{(i)}(t)=-2\bm{b}_{\bm{k}}^{(i)}\times \bm{S}_{\bm{k}}^{(i)}(t)$,
where $i=\pm$ and $\bm{S}_{\bm{k}}^{(i)}(t)\equiv \left<\hat{\bm{S}}_{\bm{k}}^{(i)}\right>$ are the expectation value of Anderson pseudo-spin operators. The time dependent gap can be written using pseudo-spins as: $
\Delta(t)=\frac{U}{2N_c}\sum_{\bm{k},i=\pm}\left(S_{\bm{k}}^{(i)x}+
\mathrm{i}S_{\bm{k}}^{(i)y}\right)$.

\begin{figure}[htbp]
\centering
\includegraphics[scale=0.18]{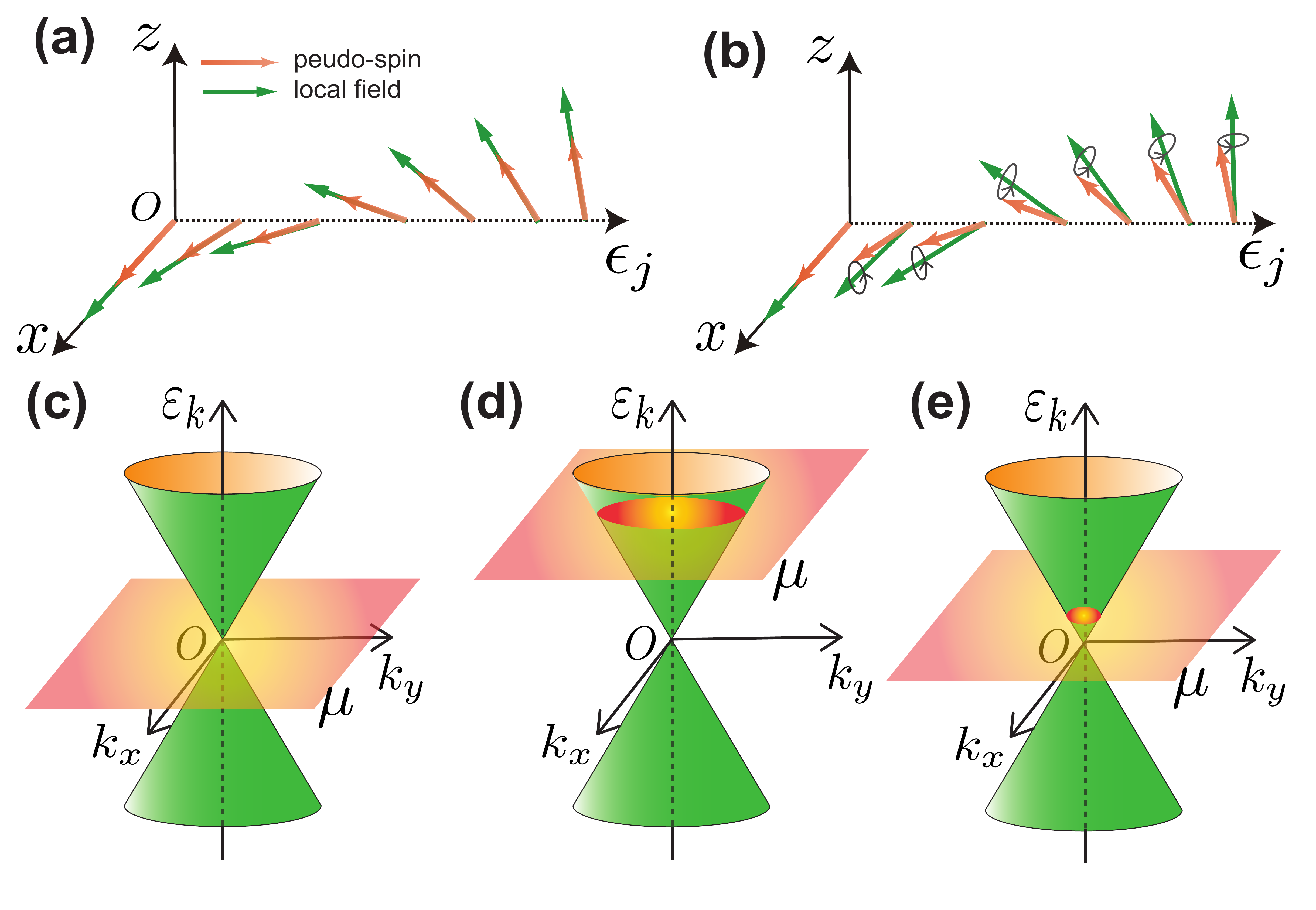}
\caption{(Color online) Quenched dynamics illustration and three doping cases for honeycomb lattice. {\bf(a)} When $t\le 0$, the system is in the BCS ground state, the pseudo-spins align in the direction of their local fields. {\bf(b)} At $t=0^+$, we change the interaction strength abruptly to make the system out of equilibrium. The pseudo-spins start to precess around their local fields, while the local fields also change due to their dependence on pseudo-spins. {\bf(c)} The half filling case: $\mu=0$, where $\varepsilon_{\bm{k}}\equiv\pm |\gamma_{\bm{k}}|$. {\bf(d)} The high doping limit: $\mu\gg \Delta_{0f}$. {\bf(e)} The low doping limit: $\mu \sim \Delta_{0f}$.
}\label{fig:pseudospin}
\end{figure}

For simplicity, we can also label the pseudo-spins by energy state $\epsilon_j$ rather than $\bf{k}$, so that we can combine the two sets of pseudo-spins as a single set. Explicitly, the equations of motion and time dependent gap can be rewritten as:

\begin{align}
&\frac{\partial}{\partial t} \bm{S}_j(t)=-2\bm{b}_j(t)\times \bm{S}_j(t)\label{eom2}\\
&\Delta(t)=\frac{U}{2N_c}\sum_{j}\left(S_j^{x}(t)+
\mathrm{i}S_j^{y}(t)\right)\label{Delta-spin2}
\end{align}
with:
\begin{equation}
 \bm{b}_j(t)=(\Delta^R(t), \Delta^I(t), \epsilon_j)\label{field-def}
\end{equation}
where $\epsilon_j\in(-\omega_D, \omega_D)$, and $\bm{S}_j$ can be view as the classical spin with length $\frac{1}{2}$.  Writing like this, the additional DOS information is needed. It satisfies $D(\epsilon)\propto |\epsilon-\mu|$, for we have a 2D linear dispersion near the Dirac point before superconducting, see [FIG.(\ref{fig:pseudospin}(c))].

The quenched dynamics is as follows: at $t\le 0$, the system is in equilibrium with the initial interacting strength $U_i$. From the spin Hamiltonian, the initial spins are parallel to their local fields [Fig. \ref{fig:pseudospin}(a)].  At $t=0^+$, we change the interaction strength to $U_f$, then the local fields change immediately for the sudden change of $\Delta(t)$. Therefore, the current spin configuration is no longer stable. According to equation (\ref{eom2}), they will precess around their local fields[Fig.\ref{fig:pseudospin}(b)], which in turn will change the gap and the local fields simultaneously by equation (\ref{Delta-spin2}) and (\ref{field-def}) . We denote  $\Delta_{0i}$ and $\Delta_{0f}$ as the corresponding equilibrium gap when the interaction strength are $U_i$ and $U_f$, respectively. In the following, they are used to describe the quenched dynamics for convenience.
 
\section{three doping cases for honeycomb lattice}
We consider the dynamics of three doping cases for honeycomb lattice as shown in Fig. \ref{fig:pseudospin}(c, d, e): half filling, high doping limit and low doping limit.
\subsection{Half filling}
Without loss of generality, we choose the initial gap $\Delta_{0i}$ to be real. The particle-hole symmetry guarantees the gap to be real throughout the evolution\cite{Levitov2006}. The problem is to solve a system of coupled differential equations (\ref{eom2}) with the initial condition: $\bm{S}_{j}(0)=\left( \frac{\Delta_{0i}}{2\sqrt{\Delta_{0i}^2+\epsilon_{j}^2}},\,0,\,\frac{\epsilon_j}{2\sqrt{\Delta_{0i}^2+\epsilon_{j}^2}}\right)$, 
where in this case the gap and local fields are related with the pseudo-spins as: $\Delta(t)=\frac{U_f}{2N_c}\sum_{j}S_j^x(t)$ and $\bm{b}_{j}(t)=\left( \Delta(t),\,0,\,\epsilon_j\right)$.
The DOS in the half filling case is proportional to  $|\epsilon|$.

\begin{figure}[htbp]
 \hspace*{-20pt}
 \noindent\includegraphics[scale=0.23]{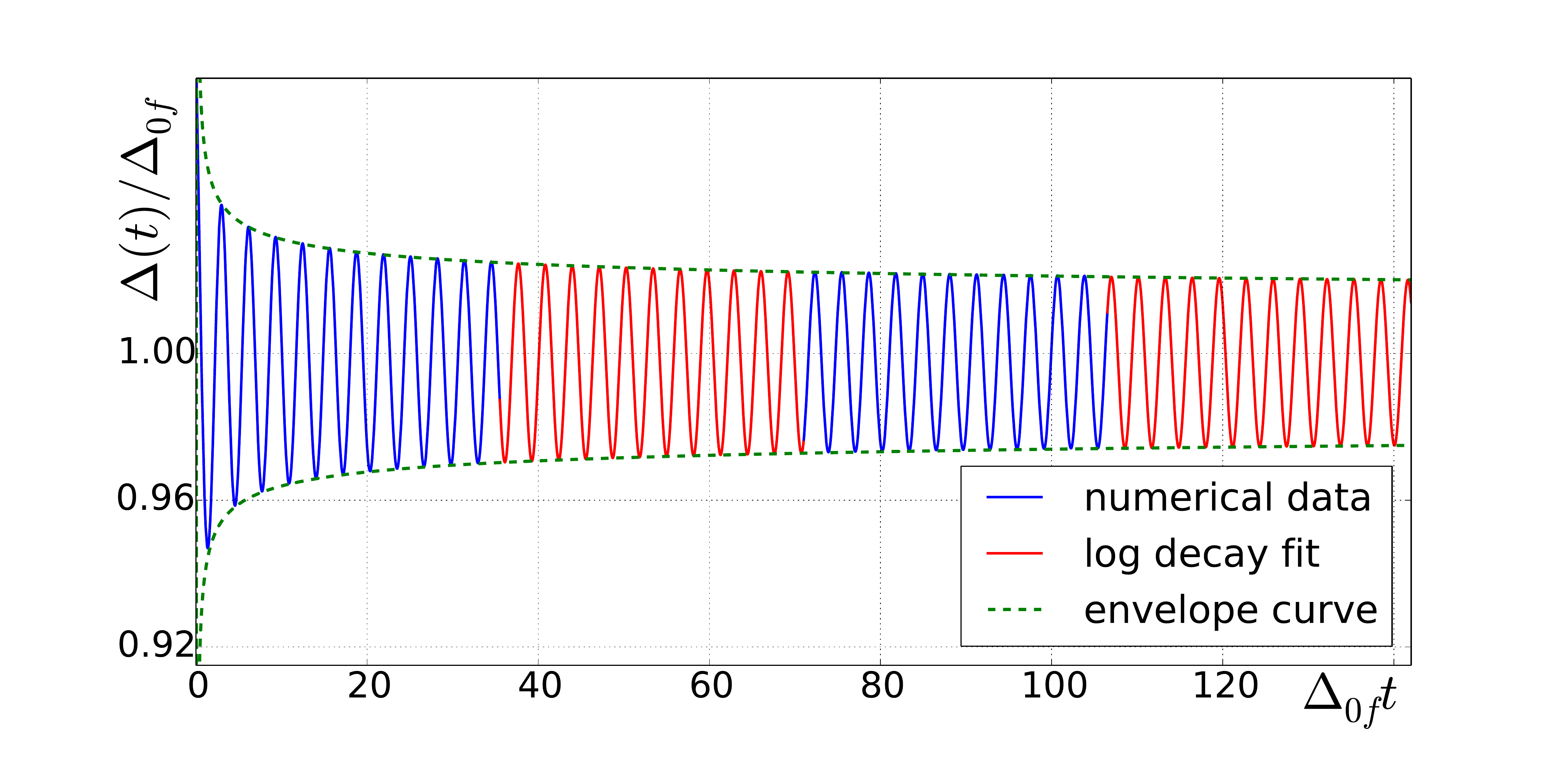}
 \caption{(color online) Half filling. The numerical data  (blue) obtained from simulating $N=50000$ energy levels for $\Delta_{0i}=0.013$ and $\Delta_{0f}=0.012$, with Deybe energy $\omega_D=0.5$. The red curve is the fit by equation (\ref{log-fit}), while the green dotted line are the envelope curves.}\label{fig:half-filling}
 \end{figure}

We numerically simulate equation (\ref{eom2}) with $N=50000$ energy levels and the Debye cut-off energy $\omega_D=0.5$. The method we use is the Runge-Kutta
of the 8-th order with an adjustable time step to meet a sufficient high precision. Other numbers of energy levels are also tried to verify that the results are unaffected by the finite size effect. We also adopt the weak coupling limit($\Delta_{0f}\ll \omega_D$) and the weak quench limit($\delta\Delta_0\equiv\Delta_{0i}-
\Delta_{0f}\ll\Delta_{0f}$). To satisfy this, we quench from $\Delta_{0i}=0.013$ to $\Delta_{0f}=0.012$. The result is shown in FIG.(\ref{fig:half-filling}): the data is well fitted by a log-decay function: 
\begin{equation}\label{log-fit}
\frac{\Delta(t)}{\Delta_{0f}}=a+\frac{2b\delta\Delta_0}{\Delta_{0f}}
\frac{\cos(c\Delta_{0f}t+d)}{\ln(e\Delta_{0f}t)}
\end{equation}
The envelope functions $a\pm 2b\delta\Delta_0/\Delta_{0f}\ln(e\Delta_{0f}t)$ are used for indicating the log-decay behaviour.

The fitted parameter
are: $a=0.9975, b=1.091, c=1.994, d=0.2554, e=22.36$. We find that $c=2a$ is almost exactly satisfied, which means that $\Delta(t)$ oscillates with the $2\Delta_\infty$ angular frequency, indicating it is the Higgs amplitude mode. However, the mode has a logarithmic decaying property in the present case, while it decays as $1/\sqrt{t}$ in the normal superconductors. This slow decaying behaviour suggests the Higgs mode in the half-filling superconducting honeycomb lattice has a much longer lifetime than that in the usual superconductors\cite{Tsuchiya2013}. We also note that $a$ is slightly smaller than $1$, meaning $\Delta_\infty<\Delta_{0f}$. Explicitly, we find $1-a\approx \delta\Delta_0^2/3\Delta_{0f}^2$. The similar behaviour has been pointed out in the previous literature for the normal superconductors, claiming that the difference is of order $\delta\Delta_0^2/6\Delta_{0f}^2$\cite{Yuzbashyan2006Linear,Yuzbashyan2015}.

\begin{figure}[htbp]
 \hspace*{0pt}
 \noindent\includegraphics[scale=0.41]{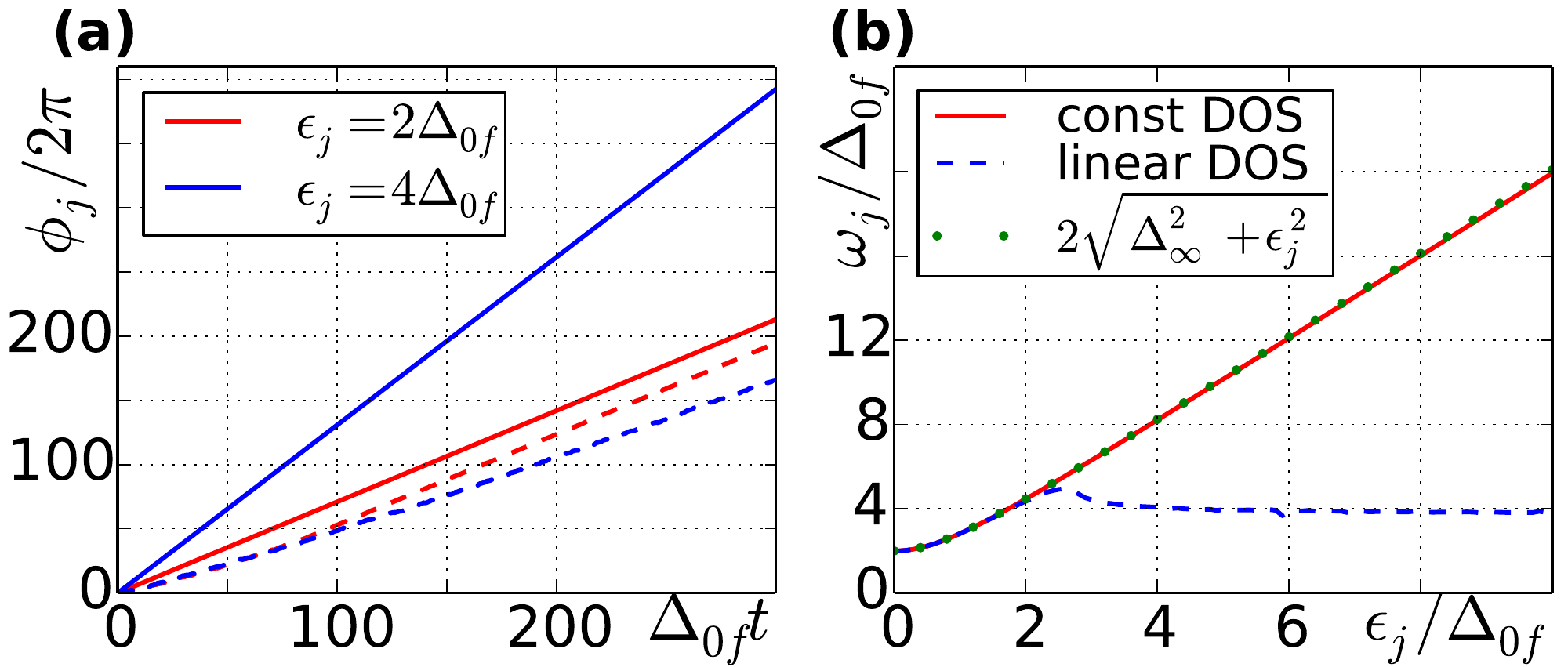}
 \caption{(color online) Phase dynamics for $\Delta_{0i}=0.013$ and $\Delta_{0f}=0.012$ at half filling. The solid lines are for $D(\epsilon)=1$, while the dashed lines are for $D(\epsilon)\propto |\epsilon|$.  $\bf{(a).}$ The precession phases $\phi_j$ for $\epsilon_j=2\Delta_{0f},4\Delta_{0f}$. They are almost linear for the large time dynamics and $\phi_j$ for constant DOS has larger
 ``phase slope". $\bf{(b).}$ The time averaged precession frequency $\omega_j$. For constant DOS, $\omega_j$ coincides with quasiparticle energy spectrum. For linear DOS case, the flatter $\omega_j$'s dispersion gives rise to in a weaker dephasing, therefore a slower decay of the amplitude.}\label{fig:phi-omega}
 \end{figure}

The slower decaying property compared with normal  superconductors can be qualitatively understood by studying the phase dynamics of the single pseudo-spins on different energy levels\cite{Levitov2006}. Explicitly, we numerically calculate the precession angle $\phi_j(t)$ of pseudo-spin $\bm{S}_j$ around the time independent vector 
$\bm{b}_j^\infty\equiv(\Delta_{\infty},0,\epsilon_j)$. As shown in FIG.\ref{fig:phi-omega}(a), in the long time limit, the phase become linear
with respect to time so that we can character the precession frequency by the time averaged frequency $\omega_j=\left<\omega_j(t)\right>=[\phi_j(t_{max})
-\phi_j(0)]/t_{max}$. In FIG.\ref{fig:phi-omega}(b), we compare $\omega_j$ for constant and linear DOS. For constant DOS, $\omega_j$ is equal to the quasi-particle spectrum  $2\sqrt{\Delta_\infty^2+\epsilon_j^2}$. In the region when $\epsilon_j\lesssim 2\Delta_{0f}$, $\omega_j$ for both cases coincide with each other. However 
in the higher energy region,  $\omega_j$ for linear DOS is much flatter than that for constant DOS. The decaying of
the amplitude is due to the dephasing mechanism for the precession of pseudo-spins. The flatter dispersion of $\omega_j$ represents a more synchronized precession of the pseudo-spins, resulting in a slower decaying of the amplitude.

To quantitatively understand the fitting equation (\ref{log-fit}), we solve equations of motion (\ref{eom2}) by linearizing it around $\bm{S}_j^f\equiv \left( \frac{\Delta_{0f}}{2\sqrt{\Delta_{0f}^2+\epsilon_{j}^2}},\,0,\,\frac{\epsilon_j}{2\sqrt{\Delta_{0f}^2+\epsilon_{j}^2}}\right)$ and $\bm{b}_j^f\equiv \left( \Delta_{0f},\,0,\,\epsilon_j\right)$:

\begin{align}
\frac{\partial}{\partial t} \delta S_{j}^x (t)&= 2\epsilon_{j} \delta S_{j}^y(t) \nonumber
\\
\frac{\partial}{\partial t} \delta S_{j}^y(t) &= \frac{\epsilon_{j} }{\sqrt{\Delta_{0f}^2+\epsilon_{j}^2}} \delta\Delta(t)+2\Delta_{0f}\delta S_{j}^z(t)-2\epsilon_{j} \delta S_{j}^x(t) \nonumber
\\
\frac{\partial}{\partial t} \delta S_{j}^z(t) &= -2\Delta_{0f} \delta S_{j}^y(t) \label{linear-eqn1}
\end{align}
where $\delta \Delta(t) \equiv\Delta(t)-\Delta_{0f}$ and $\delta \bm{S}_{j}(t)\equiv \bm{S}_{j}(t)-\bm{S}_{j}^f$.
The above coupled differential equation can be solved by Laplace transform: $\mathcal{L}\left[f(t)\right]\to \bar{f}(s)$. %
In the thermodynamic and the weak coupling limit, we arrive at the final form of $\overbar{\delta \Delta}(s)$:
\begin{equation}\label{eqn:image-func}
\overbar{\delta\Delta}(s)=\frac{\delta\Delta_0}{2\Delta_{0f}}\left(\frac{1}{\left(\frac{s}{2\Delta_{0f}}\right)}-\frac{1}{\left[\left(\frac{s}{2\Delta_{0f}}\right)^2+1\right]\tan^{-1}\left(\frac{s}{2\Delta_{0f}}\right)} \right)
\end{equation}

By inverse Laplace transform, we can get the approximate form of $\Delta(t)$ (see Appendix \ref{appendixA}):

\begin{equation}\label{eqn:asymptotic}
\Delta(t)\approx \Delta_f+2\delta\Delta_0\frac{\cos 2\Delta_f t}{\ln 4\Delta_f t}
\end{equation}

\subsection{Doping cases}
In the high doping limit ($\mu \gg \Delta_{0f}$) as illustrated in Fig. \ref{fig:pseudospin}(d). The system without attractive interaction is basically a normal metal, therefore we expect the Higgs mode will have the square-root decaying behaviour. To verify this, we choose $\mu=0.12=10\Delta_{0f}$ and simulate equation (\ref{eom2})-(\ref{field-def}) with other parameters  equal to those in the half-filling case. The result is shown 
in Appendix \ref{appendixB}. We can see $|\Delta(t)|$ indeed decays as $1/\sqrt{t}$, with the oscillation frequency equals to $2\Delta_\infty$.
 
\begin{figure}[htbp]
 \hspace*{-15pt}
 \noindent\includegraphics[scale=0.40]{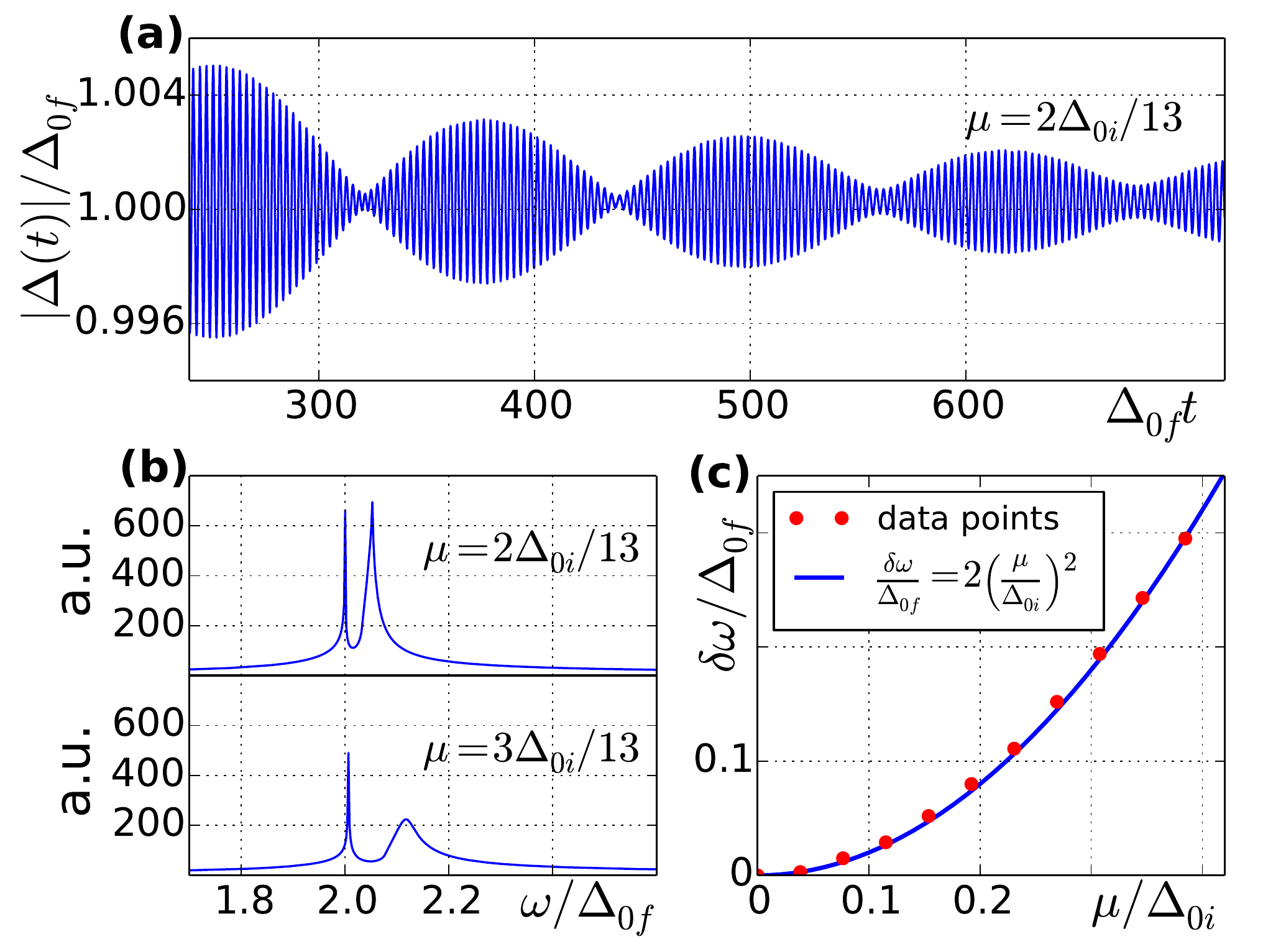}
 \caption{(color online) Low doping case. The quench parameters are the same as in FIG.\ref{fig:half-filling}. {\bf(a)} The two slightly different frequencies give rise to a beat pattern of the amplitude mode. {\bf(b)} The frequencies obtained by discrete fourier transfrom (DFT) of $|\Delta(t)|$. Both frequencies increases as $\mu$ increases, while the larger one increase more noticeably. Besides, the larger frequency peak
also broadens and will eventually disappear as $\mu$ increases, accomplishing the gradual transform from logarithimic decay to square root decay. {\bf(c)} The frequencies data (red dots) collect by DFT of different values of $\mu$, they fit quite well by the empirical formula (blue line). }\label{fig:low-dp}
 \end{figure}

To see how the mode change from 
the logarithmic decay to the $1/\sqrt{t}$ decay, we 
investigate the low doping limit where $\mu \sim \Delta_{0f}$ [Fig. \ref{fig:pseudospin}(e)]. By simulating equation (\ref{eom2})-(\ref{field-def}) with several different values of $\mu$, we find there are two frequencies in the low doping case: one is the Higgs frequency $2\Delta_\infty$, the other is slightly larger than the first one, resulting in a beat pattern as shown in FIG.\ref{fig:low-dp} (a). As $\mu$ increases, we find both 
frequencies increase. However, the Higgs frequency increases only slightly, while the lager frequency increases more remarkably and the peak
broadens[Fig.\ref{fig:low-dp} (b)]. Physically, the decay of the Higgs mode is
due to its interaction with the bottom part of the particle-hole continuum\cite{Volkov1973,Barankov2004}. As we doped away from half filling, those states most responsible for the damping increase, resulting a faster decaying behaviour. When $\mu$ is large enough (about 2$\Delta_{0f}$), the second peak can hardly be discerned and the transform from the logarithmic decay to square-root decay accomplishes. We also find a very interesting empirical formula, associating the difference of the two frequencies $\delta\omega$ with the chemical potential $\mu$ as: $
\frac{\delta\omega}{\Delta_{0f}}=2\left(\frac{\mu}{\Delta_{0i}}\right)^2$.
  
\section{Dirac semimetal case} 
\begin{figure}[htbp]
 \hspace*{0pt}
 \noindent\includegraphics[scale=0.40]{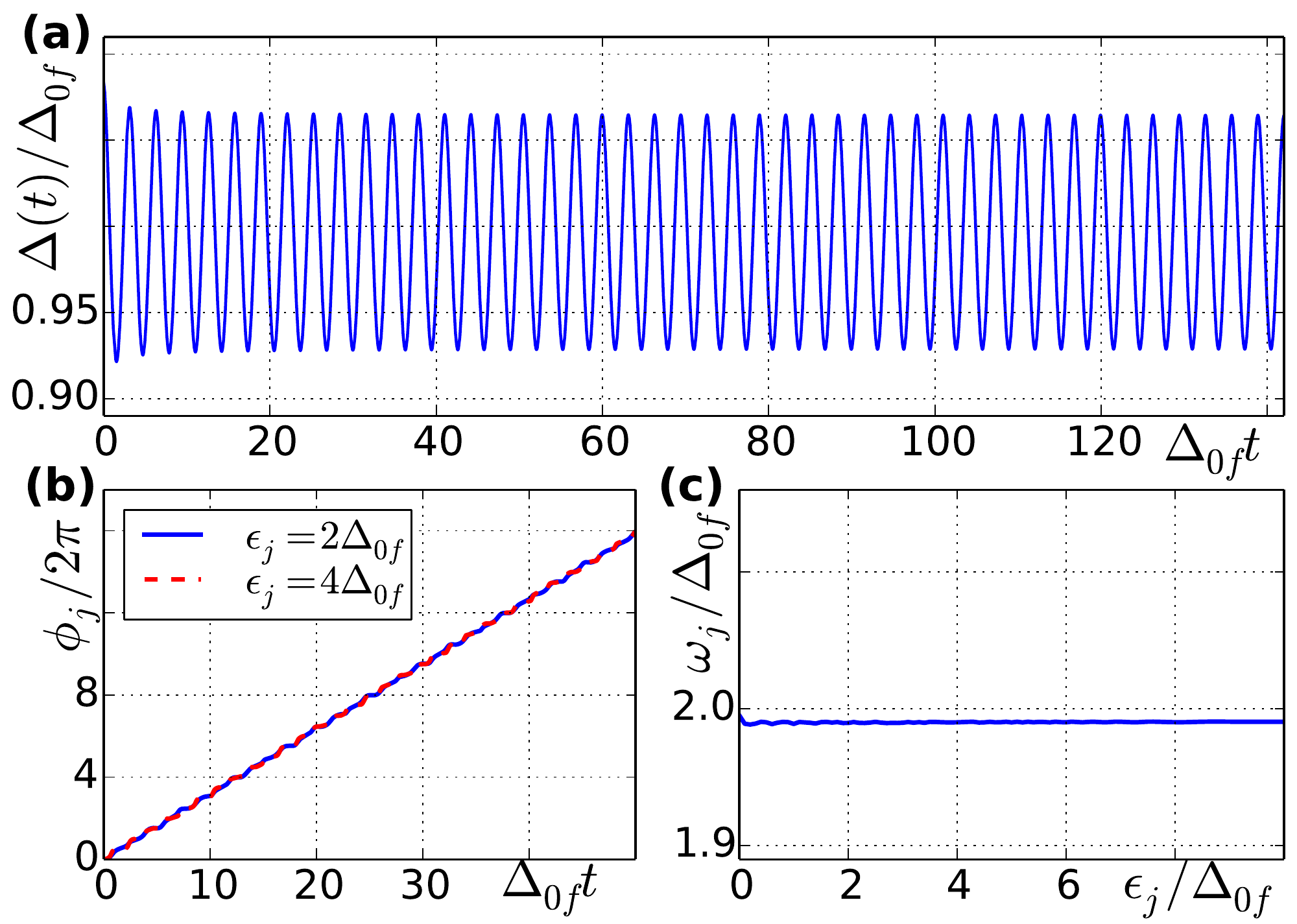}
 \caption{(color online) 3D Dirac semimetal case. The quench parameters are equal to those in FIG.\ref{fig:half-filling}.   {\bf (a).} The Higgs mode shows an undamped oscillation. {\bf (b).} Precession phase of single spin on energy levels $\epsilon_j=2\Delta_{0f}, 4\Delta_{0f}$. {\bf (c).} The precession of different pseudo-spins synchronize.}\label{fig:SM}
 \end{figure}

We extend our calculation to the 3D Dirac semimetal case. The DOS is proportional to $\epsilon^2$ when the Fermi level is on the Dirac point. We numerically solve the collective motion of pseudo-spins with all the parameters equal to those in the half filling honeycomb lattice case. We find the Higgs amplitude mode in this case exhibits an undamped oscillation as shown in FIG.\ref{fig:SM}(a). To explain this, we study the phase dynamics $\phi_j(t)$ of each pseudo-spin $\bm{S}_j(t)$ that precess around its own time independent vector $\bm{b}_j^\infty$. From  FIG.\ref{fig:SM} (b, c), we can see that all the pseudo-spins precess with the same angular frequency $2\Delta_\infty$. Therefore, for the two instances of time separated by $T=\pi/\Delta_\infty$, the whole pseudo-spins' configuration is identical. Since $\Delta(t)$ depends explicitly on the sum of $x$ component of all the pseudo-spins, it must be periodic and undamped. Compared with 2D case at half filling, the particle-hole continuum most responsible for the damping consist a even smaller fraction of the whole phase space. Therefore, the damping originating from the interaction with those states is negligible. We note that the above discussion is for the singlet pairing case. However, the triplet pairing is also possible, which has three independent Higgs mode\cite{Rosenstein2015}. Studying the time evolution of these Higgs mode would also be interesting.

\section{discussion and summary}  For the 2D superconducting Dirac fermion case, the quenched process can be realized on the two-component cold Fermi gases trapped in a honeycomb optical lattice\cite{Esslinger2012}, with an attractive Hubbard $U$ tunable by the Feshbach resonance\cite{Esslinger2010}. The Higgs mode in this case can be detected with the rf-absorbtion techniques~\cite{Dzero07,Chin04}. As for the Higgs amplitude mode in 3D case, the observation is made possible by the recent discovery of superconductivity in Dirac semimetals \cite{Wangjian2015,Aggarwal2015}, together with the development of the ultrafast THz pump-probe spectroscopy\cite{Nelson2013}.  In principle, the measurement should be similar to the already discovered Higgs mode in the clean NbN film\cite{Matsunaga2013}. One can use an intense monocycle THz pump pulse to generate the Higgs amplitude mode in the superconducting $\text{Cd}_3\text{As}_2$ thin film. Immediately after that, a probe pulse also irradiates to the sample. By measuring the pump-probe delay time and the wave form of the transmitted probe pulse, one can resolve the time evolution of the Higgs mode inside the sample\cite{Matsunaga2013,Matsunaga2014}.

In summary, we find the Higgs amplitude mode in half-filling honeycomb lattice has a logarithmic decaying behaviour. It can be understood by studying its phase dynamics, and by analytically solving the linearized equations of motion. The dynamics of doped cases in honeycomb lattice is also studied. As for the three dimensional Dirac semimetals case, we find the Higgs mode exhibits an undamped oscillation when the Fermi level is at the Dirac point.

\section{Acknowledgement}
This work was financially supported by NBRP of China (2012CB821402 and 2015CB921102) and NSF-China under Grants Nos.11534001, 11504008 and 11304280.

\appendix

\section{INVERSE LAPLACE TRANSFORM OF EQ.\ref{eqn:image-func}}\label{appendixA}
By doing the Laplace transform of the linearized equations of motion, we get the following equation for $\overbar{\delta\Delta}(s)$ up to the linear order of $\delta\Delta_0$:

\begin{align}
&\overbar{\delta\Delta}(s) \sum_{j}\frac{1}{\left(s^2+4\Delta_{0f}^2+4\epsilon_{j} ^2\right)\left(\Delta_{0f}^2+\epsilon_{j} ^2\right)^{\frac{1}{2}}}\nonumber
\\
=&\frac{s\delta\Delta_0}{s^2+4\Delta_{0f}^2}\sum_{j}\frac{\epsilon_{j} ^2}{\left(s^2+4\Delta_{0f}^2+4\epsilon_{j} ^2\right)\left(\Delta_{0f}^2+\epsilon_{j} ^2\right)^{\frac{3}{2}}}\label{sum}
\end{align}

In the thermodynamic limit and weak coupling limit, we have
$\sum_j f(\epsilon_j)\propto \int_0^{\omega_D} f(\epsilon)\epsilon\mathrm{d}\epsilon\approx \int_0^{\infty}f(\epsilon)\epsilon\mathrm{d}\epsilon$. After the integration, we get equation (\ref{eqn:image-func}) in the main text.

Using the similarity theorem  $\mathcal{L}^{-1}\left[\bar{f(s/a)}\right]=af(at)$, we need only to find the the inverse Laplace transform of $\bar{f}(s)=1/(s^2+1)\tan^{-1}s$. We achieve this by evaluating the Bromwich integral:
\begin{equation}
f(t)=\frac{1}{2\pi \mathrm{i}}\int_{\gamma-i\infty}^{\gamma+i\infty} \mathrm{d}s\bar{f}(t) e^{st}
\end{equation}
where $\gamma$ should be larger than the real part of any poles in the integrand. 
\begin{figure}[htbp]
\centering
\includegraphics[scale=0.35]{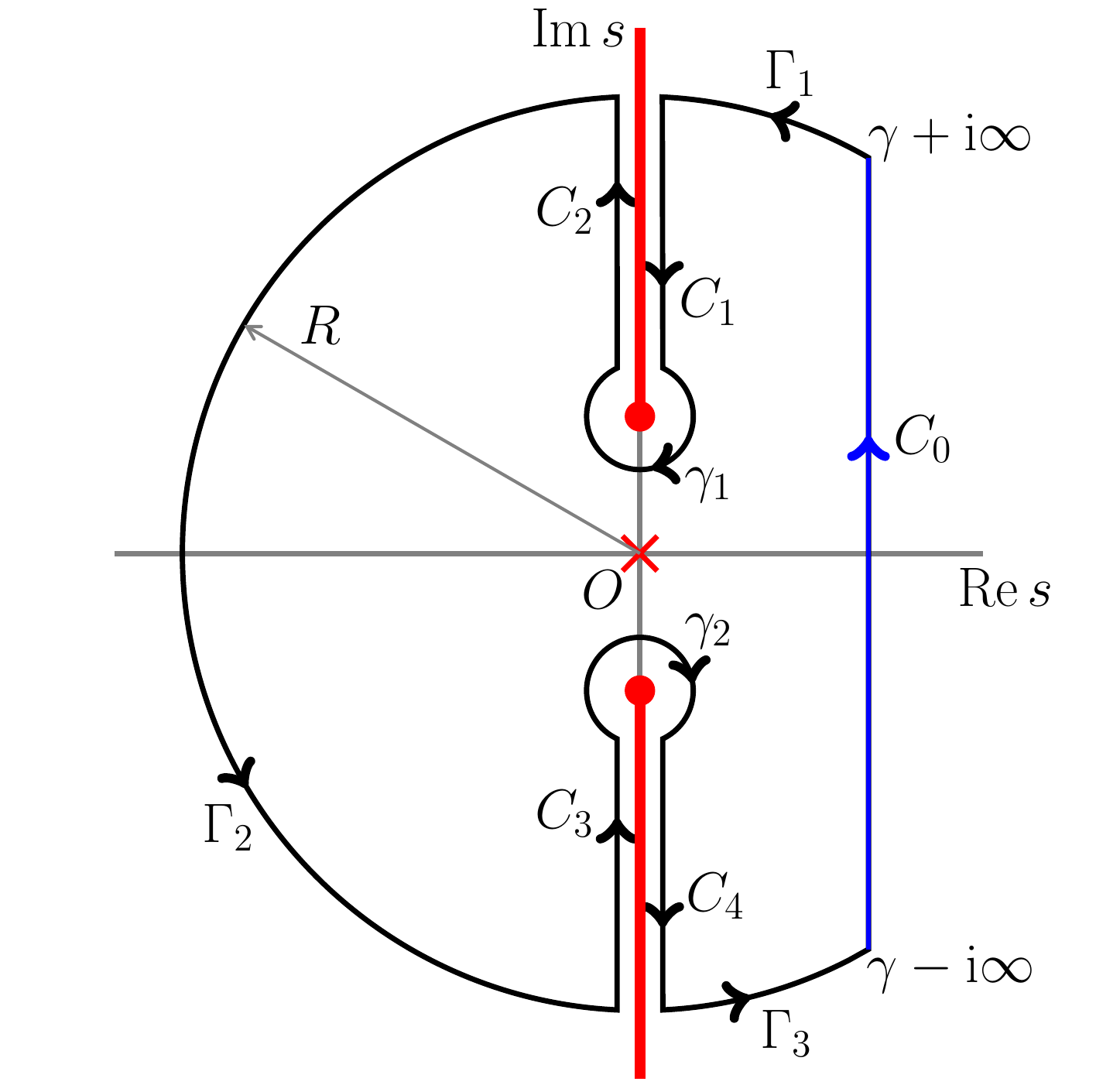}
\caption{(color online) The countour of the integral. The red cross represents the pole at $s=0$, the red points are branch points at $s=\pm \mathrm{i}$, the red lines are the two branch cuts.}\label{contour}
\end{figure}

We choose the contour shown in Fig. \ref{contour}, and use Cauchy's integral theorem to evaluate the Bromwich integral $C_0$ marking in blue. The Jordan's lemma tells us the contributions from big arcs $\Gamma_1,\Gamma_2,\Gamma_3$ are zero, and it is easy to verify that the integrals along the small arcs $\gamma_1$ and $\gamma_2$ have no contributions either. The only remaining parts are the pole at origin and line integrals $C_1$ to $C_4$. So we have:
\begin{align}
f(t)&=\theta(t)-4I_2(t)
\\
I_2(t)&=\Re\left[e^{it}\int_0^\infty \frac{e^{i xt}}{(x^2+2x)\left[\left(\ln\frac{x}{x+2}\right)^2+\pi^2\right]}\mathrm{d}x\right]\label{eqn:I2}
\end{align}

\begin{figure}[H]
\centering
\includegraphics[scale=0.42]{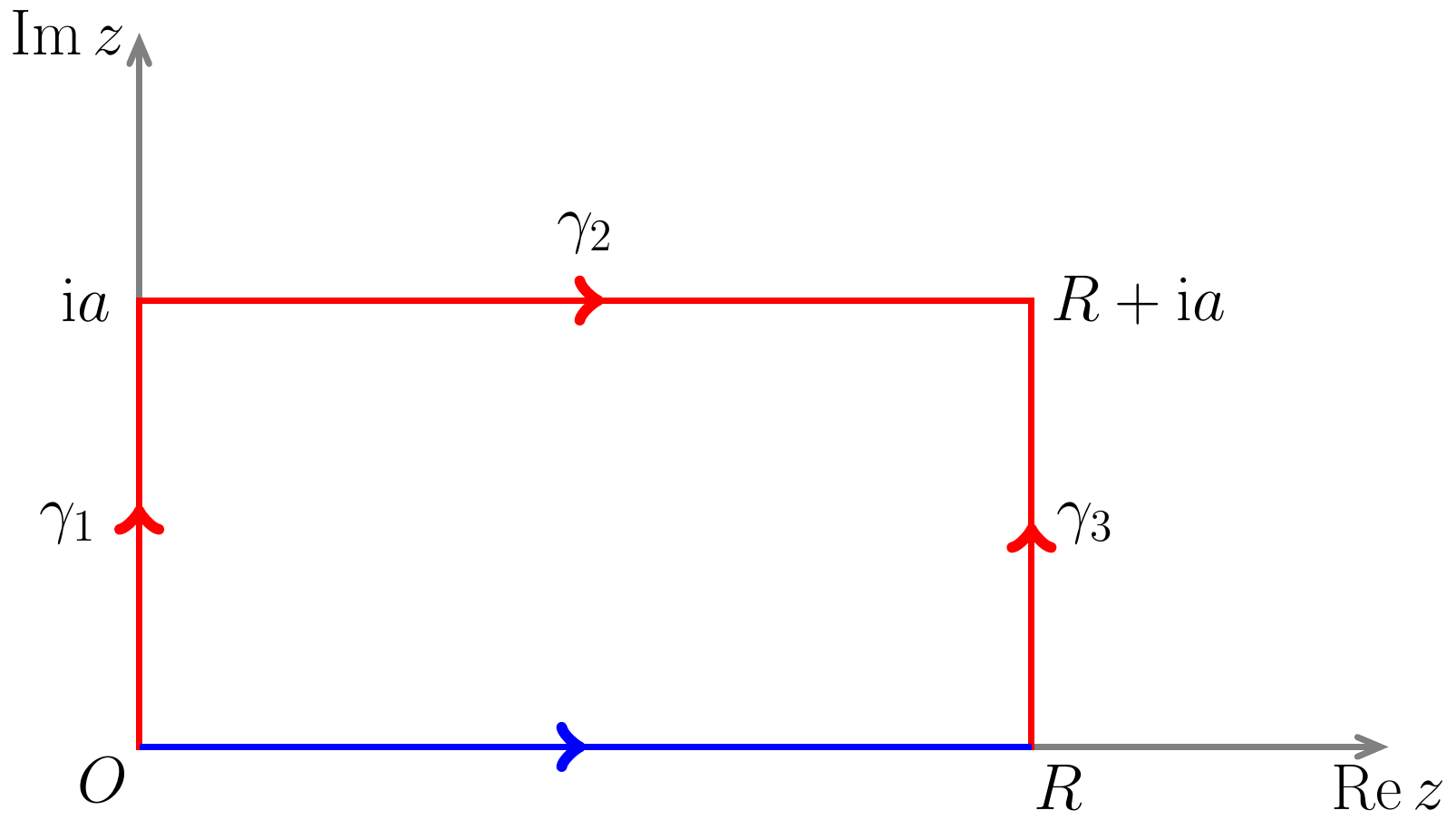}
\caption{(color online)The contour for $I_2(t)$, $a$ is a small real positive number, the integral $I_2(t)$ (blue) is replaced by the contour in red, while integration along $\gamma_2$ and $\gamma_3$ are zero.}
\label{rect-contour}
\end{figure}

We use the contour in Fig. \ref{rect-contour} to evaluate equation(\ref{eqn:I2}), and the only remaining contribution is from the line integral $\gamma_1$. To the leading order, we have:
\begin{equation}
I_2(t)=\Re\left[e^{it}\int_0^{2a} \frac{e^{-2yt}}{2y(\ln y)^2}\mathrm{d}y\right]
\end{equation}
For large enough $t$, the above integral can be conducted by using a result by A. Erdélyi\cite{Erdélyi}, thus we obtain equation(\ref{eqn:asymptotic}) in the main text.

\section{High doping limit case}\label{appendixB}
We choose $\mu=0.12$ in this case. Because the exact particle-hole symmetry is absent when $\mu\neq 0$, $\Delta(t)$ will acquire a time-depended phase during the evolution, thus we plot the amplitude $|\Delta(t)|$ in the figure. We fit the data using the following equation provided in many literatures\cite{Volkov1973,Yuzbashyan2006Linear,Tsuji2014}:

\begin{equation}\label{square-fit}
\frac{|\Delta(t)|}{\Delta_{0f}}= a + \frac{2b\delta\Delta_0}{\pi^{\frac{3}{2}}\Delta_{0f}\sqrt{\Delta_{0f} t}}\cos\left(c\Delta_{0f} t +d\frac{\pi}{4}\right)
\end{equation}

The fitting parameters are: $a=1.0050,  b=0.5142, c=2.0101,  d=0.9827$. We see $c=2a$ is almost exactly satisfied, indicating this is the Higgs amplitude mode. However, $a$ is 
slightly greater than $1$, meaning $\Delta_{\infty}$ is slightly greater than $\Delta_{0f}$. This is not so surprising because the relation $\Delta_{\infty}\approx\Delta_{0f}-
\delta\Delta_0^2/6\Delta_{0f}$ is obtained under the strictly 
constant density of state condition. In conclusion, in the high doping limit, the system behaves as a normal metal without interaction, resulting the $1/\sqrt{t}$ decaying property of the amplitude $|\Delta(t)|$.
\begin{figure}[H]
 \hspace*{0pt}
 \noindent\includegraphics[scale=0.21]{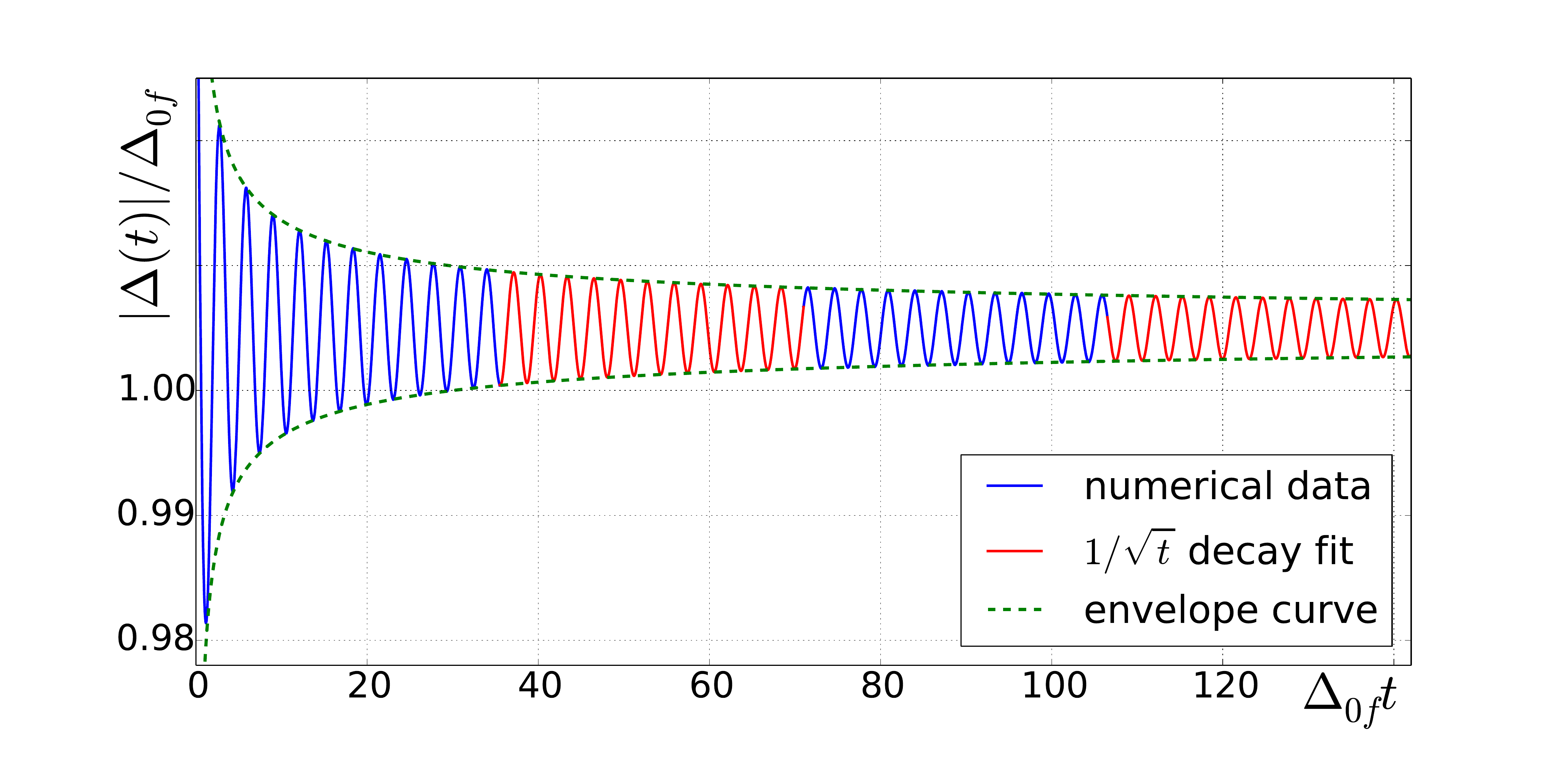}
 \caption{(color online) High doping limit with $\mu=0.12$, other parameters are same as those in FIG. 2 in the main text. The numerical data (blue) is well fitted by equation (\ref{square-fit}).}\label{fig:high-dp}
 \end{figure}

\end{document}